\documentclass[osajnl,twocolumn,showpacs,superscriptaddress,10pt]{revtex4-1} 
\usepackage{amsmath,amssymb,graphicx}
\begin{document}

\title{Quantum atomic lithography via cross-cavity optical Stern-Gerlach Setup}
\author{C. E. M\'{a}ximo}
\author{ T. B. Batalh\~{a}o}
\author{, R. Bachelard}
\author{ G. D. de Moraes Neto}
\affiliation{Instituto de F\'{\i}sica de S\~{a}o Carlos, Universidade de S\~{a}o
Paulo, Caixa Postal 369, 13560-590 S\~{a}o Carlos, SP, Brazil }
\author{ M. A. de Ponte}
\affiliation{Universidade Regional do Cariri, Departamento de F\'{\i}sica,
BR-63010970, Juazeiro Do Norte, CE, Brazil}
\author{M. H. Y. Moussa.}
\affiliation{Instituto de F\'{\i}sica de S\~{a}o Carlos, Universidade de S\~{a}o
Paulo, Caixa Postal 369, 13560-590 S\~{a}o Carlos, SP, Brazil }
\begin{abstract}
We present a fully quantum scheme to perform 2D atomic lithography based on a
cross-cavity optical Stern-Gerlach setup: an array of two mutually orthogonal
cavities crossed by an atomic beam perpendicular to their optical axes, which
is made to interact with two identical modes. After deriving an analytical
solution for the atomic momentum distribution, we introduce a protocol
allowing us to control the atomic deflection by manipulating the amplitudes
and phases of the cavity field states. Our quantum scheme provides
subwavelength resolution in the nanometer scale for the microwaves regime.

\end{abstract}

\maketitle

At the beginning of the 1990s, the optical Stern-Gerlach (OSG) effect
\cite{Sleator} was explored in a number of studies \cite{HAS,FH,BB}, with a
view to extracting information about a cavity field state through its
interaction with an atomic meter. Relying on the fact that the momentum
distribution of scattered atoms follows the photon statistics of the field
state, strategies have been devised to reconstruct the statistics \cite{HAS}
and even the full state of a cavity mode \cite{FH}. These OSG strategies
differ from other measurement devices in quantum optics, such as quantum
nondemolition \cite{Brag} and homodyne techniques \cite{VR}, that have been
extensively explored from the 1990s until now \cite{Exp}. More recently, a
cross-cavity OSG has been proposed ---where a beam of atoms is made to cross
two orthogonal cavities--- to measure the location and center-of-mass
wave-function of the atoms \cite{Zubairy, ADK}. Although the cross-cavity OSG
has not yet been implemented experimentally, the cross-cavity setup has been
built to test Lorentz invariance at the $10^{-17}$ level \cite{Peters}.

In addition to the developments in probing atomic and cavity-field states,
atomic lithography ---where classical light is used to focus matter on the
nanometer scale--- has also witnessed considerable progress in recent decades
\cite{Pfau}. The atom-light interaction is manipulated to assemble structured
array of atoms with potential applications to nanotechnology-related fields.
Beyond the achievements in the growth of spatially periodic and quasi-periodic
\cite{Kampen} atomic patterns \cite{Pfau}, recent works have explored the
possibility of creating nonperiodic arrays by using complex optical fields
\cite{12-13,Saffman}.

In this paper, we present a scheme to realize two-dimensional (2D) quantum
atomic lithography. In order to characterize it, we derive an analytical
solution for the 2D OSG problem. We consider the cross-cavity OSG setup
sketched in Fig. 1, where, before entering the cavities, the atoms are
confined by a circular pinhole to a small region of space, centered around the
superimposed nodes of the two cavity modes. Differently from the developments
in Refs. \cite{Zubairy, ADK}, where dispersive atom-field interactions take
place, we assume the two-level atoms to undergo simultaneous and resonant
interactions with two identical modes, one from each cavity, thus being
deflected in the plane defined by the two mutually perpendicular cavities'
optical axes. An appropriate ansatz on the spatial distribution of the atoms
across the pinhole enables us to derive an analytical expression for the
atomic momentum distribution after the atom-field interactions. Our protocol
to generate 2D nonperiodic complex atomic patterns is based on a map that
relates the transverse momentum acquired by the atoms to the previously
prepared cavity-field state. Interestingly, we find that the (abstract)
momentum-quadrature components of the field states are directly associated
with the (real) atomic momentum components.

Before addressing the cross-cavity OSG, it is worth mentioning previous works
in the Literature on quantized light lenses for atomic waves. We start with
the proposals for focusing and deflecting an atomic beam through quantized
field, which also addresse the process of creating regular structures with a
period of atomic size \cite{Averbukh, Rohwedder, QOPS}. There is also the
quantum prism proposal, where the deflection of an atom de Broglie wave at a
cavity mode can produce an entangled state in which discernable atomic beams
are entangled to photon Fock states \cite{Domokos}. Optical lenses made of
classical field have also been extensively studied \cite{Ole}. In a sense, we
are thus presenting a generalization of these results to perform 2D quantum
atomic lithography. Indeed we are deriving an analytical solution for the
atomic momentum distribution and introducing a protocol allowing to control
the atomic deflection through the amplitudes and phases of the cavity field
states. As it becomes clear below, a new ingredient introduced in our
developments is the use of squeezed states of the radiation fields in the
cross-cavity device to increase the resolution of the atomic momentum distribution.

In the cross-cavity OSG, sketched in Fig. 1, the beam of two-level atoms (of
transition frequency $\omega_{0}$) crosses the two cavities in a direction
perpendicular to their orthogonal optical axes, to interact resonantly with
two identical modes (of frequency $\omega=ck=2\pi c/\lambda$). To simplify the
mathematical working, we proceed to a set of reasonable approximations,
starting by assuming that both cavity modes have the same electric field per
photon ($\mathcal{E}_{0}$), thus giving rise to the same interacting dipole
moment $\mu=\mu_{x}=\mu_{y}$. We next assume that the atomic longitudinal
kinetic energy $P_{z}^{2}/2M$, being considerably higher than the typical
atom-field coupling energy $\sqrt{n}\mu\mathcal{E}_{0}$, remains practically
unaffected during the atom-field interaction time. Moreover, we also neglect
the change in the atomic transverse kinetic energy under the Raman-Nath
regime, where $\left(  \Delta P_{x}^{2}+\Delta P_{y}^{2}\right)  /2M\ll
\sqrt{n}\mu\mathcal{E}_{0}$. Finally, we proceed to the Stern-Gerlach regime
by assuming that a small circular aperture is placed in front of the array of
cavities to collimate the atomic beam in a the small region $\Delta
r\ll\lambda$ centered on the nodes of the standing-wave fields at $r=0$, thus
allowing the linearization of the usual cavity standing-wave profile: $\sin
kx\approx kx$ and $\sin ky\approx ky$. Under these assumptions, the
Hamiltonian governing the interaction of the atom at position $\left(
x=r\cos\theta,y=r\sin\theta\right)  $ with the cavity field reads
\begin{widetext}\begin{equation}
H=-\mu\mathcal{E}_{0}kr\left[  \sigma_{+}\left(  \cos\theta\text{ }%
a+\sin\theta\text{ }b\right)  +\sigma_{-}\left(  \cos\theta\text{ }a^{\dagger
}+\sin\theta\text{ }b^{\dagger}\right)  \right]  \text{,} \label{1}%
\end{equation}\end{widetext}
where $a$ and $b$ ($a^{\dagger}$ and $b^{\dagger}$) stand for the annihilation
(creation) operators of the cavity modes with optical axes in the $x$ and $y$
directions, respectively, while $\sigma_{+}=\left\vert e\right\rangle
\left\langle g\right\vert $ and $\sigma_{-}=\left\vert g\right\rangle
\left\langle e\right\vert $ describe the raising and lowering operators for
the atomic transitions. Before entering the cavities, the two-level atoms
(ground $g$ and excited $e$ states) are prepared, in a Ramsey zone, in the
superposition state $c_{g}\left\vert g\right\rangle +c_{e}\left\vert
e\right\rangle $, such that the de Broglie atomic wave packet crossing the
cross-cavity array is given by $\left\vert \psi_{atom}\right\rangle
={\int\nolimits_{0}^{\infty}}{\int\nolimits_{0}^{2\pi}}drd\theta
rf(r,\theta)\left\vert r,\theta\right\rangle \left(  c_{g}\left\vert
g\right\rangle +c_{e}\left\vert e\right\rangle \right)  $, where $\left\vert
f(r,\theta)\right\vert ^{2}$ accounts for the initial spatial distribution of
the atoms normal to the beam, as determined by the pinhole.\ Regarding the
cavity modes, we assume that they are initially prepared in the state
$\left\vert \psi_{field}\right\rangle =\sum_{m.n=0}^{\infty}\mathcal{C}%
_{m,n}\left\vert m,n\right\rangle _{ab}$. Instead of computing the spatial distribution of the atoms just after
interacting with the cavity modes at $t=\tau$, we compute, as in Ref.
\cite{FH}, the probability distribution in momentum space using the
time-of-flight technique. Since the atoms evolve as free particles for
$t>\tau$, the desired spatial distribution is simply a picture of their
momentum distribution at $t=\tau$, provided the distance traveled at $t>\tau$
is much larger than the atomic beam size. At this time, given the atom-field
entanglement in momentum space, we derive the system density matrix which,
traced over the Fock states and the internal degrees of freedom of the atoms,
leaves us with the atomic momentum distribution
\begin{widetext}\begin{align}
W\left(  \wp,\phi,\tau\right)   &  =\left\vert c_{g}\right\vert ^{2}%
{\textstyle\sum\limits_{N=0}^{\infty}}
\left\vert
{\textstyle\sum\limits_{m=0}^{N}}
\mathcal{C}_{m,N-m}\mathcal{F}_{m,0}^{g\left(  N\right)  }\right\vert
^{2}\nonumber\\
&  +\frac{1}{2}%
{\textstyle\sum\limits_{N=1}^{\infty}}
{\textstyle\sum\limits_{n=1}^{N}}
\left\vert c_{g}%
{\textstyle\sum\limits_{m=0}^{N}}
\mathcal{C}_{m,N-m}\mathcal{F}_{m,n}^{g\left(  N\right)  }+c_{e}%
{\textstyle\sum\limits_{m=1}^{N}}
\mathcal{C}_{m-1,N-m}\mathcal{F}_{m,n}^{e\left(  N\right)  }\right\vert
^{2}\nonumber\\
&  +\frac{1}{2}%
{\textstyle\sum\limits_{N=1}^{\infty}}
{\textstyle\sum\limits_{n=1}^{N}}
\left\vert c_{g}%
{\textstyle\sum\limits_{m=0}^{N}}
\mathcal{C}_{m,N-m}\mathcal{F}_{m,n}^{g\left(  N\right)  \ast}-c_{e}%
{\textstyle\sum\limits_{m=1}^{N}}
\mathcal{C}_{m-1,N-m}\mathcal{F}_{m,n}^{e\left(  N\right)  \ast}\right\vert
^{2}, \label{2}%
\end{align}\end{widetext}
where $N$ corresponds to the total number of excitations of a given subspace,
$\wp=p/\hbar k$ to the scaled atomic momentum, with $p_{x}=p\cos\phi$,
$p_{y}=p\sin\phi$. The Fourier transforms of the spatial function $f\left(
r,\theta\right)  $ reads:%

\begin{widetext}\begin{equation}
\mathcal{F}_{m,n}^{\varepsilon(N)}(\wp,\phi,\tau)=\int_{0}^{\infty}\int
_{0}^{2\pi}\frac{d\theta d\rho}{2\pi k}\rho f\left(  \frac{\rho}{k}%
,\theta\right)  \mathcal{B}_{m-\delta_{\varepsilon e},n-\delta_{\varepsilon
e}}^{(N-\delta_{\varepsilon e})}\left(  \theta\right)  \operatorname*{e}%
\nolimits^{-i\rho\left[  \wp\cos(\theta-\phi)-\sqrt{n}\Lambda\right]  },
\label{3}%
\end{equation}\end{widetext}
with $\rho=kr$, $\varepsilon$ standing for the atomic states $g$ or $e$,
$\delta_{\varepsilon e}$ for the Kronecker delta ($\delta_{ee}=1$,
$\delta_{ge}=0$), and $\Lambda=\mu\mathcal{E}_{0}\tau/\hbar$ for the
atom-field interaction parameter. Finally, the functions
\begin{widetext}\begin{subequations}
\label{4}%
\begin{align}
\mathcal{B}_{m,n}^{(N)}(\theta)  &  =\sum_{\ell=\max(0,m+n-N)}^{\min
(n,m)}\overline{\mathcal{B}}_{m,n,\ell}^{(N)}\left(  \cos\theta\right)
^{N-m-n+2\ell} \left(  \sin\theta\right)  ^{m+n-2\ell},\label{4a}\\
\overline{\mathcal{B}}_{m,n,\ell}^{(N)}  &  =\frac{(-1)^{m-\ell}%
\sqrt{m!n!\left(  N-m\right)  !\left(  N-n\right)  !}}{\ell!(m-\ell)!\left(
n-\ell\right)  !(N-m-n+\ell)!}\text{,} \label{4b}%
\end{align}
\end{subequations}\end{widetext}
follow from the Bogoliubov transform used to diagonalize Hamiltonian (\ref{1}).

In order to generate the 2D momentum distribution, we have to solve the
Fourier integrals in Eq. (\ref{3}). To this end we assume, instead of the
usual Gaussian profile, the exponential azimuthal spatial distribution of the
atoms%

\begin{equation}
f\left(  \frac{\rho}{k},\theta\right)  =\frac{1}{\sqrt{2\pi}\Delta r}%
\exp\left(  -\frac{\rho}{2k\Delta r}\right)  \text{,} \label{5}%
\end{equation}
since it enables analytical solutions to the Fourier integrals. Inserting Eq.
(\ref{5}) into Eq. (\ref{3}), we obtain%
\begin{widetext}\begin{equation}
\mathcal{F}_{m,n}^{\varepsilon(N)}\left(  \wp,\phi,\tau\right)  =\sum
_{\ell=\max(0,m+n-N-\delta_{\varepsilon e})}^{\min(m-\delta_{\varepsilon
e},n-\delta_{\varepsilon e})}\sum_{s=0}^{N-u+\delta_{\varepsilon e}}\sum
_{t=0}^{u-2\delta_{\varepsilon e}}\left(  ie^{i\phi}\right)  ^{v+\delta
_{\varepsilon e}}\mathcal{R}_{m,n,\ell,s,t,u}^{\varepsilon(N)}\mathcal{S}%
_{n,s,t}^{\varepsilon\left(  N\right)  }\text{,} \label{6}%
\end{equation}
where we have used the Newton binomial coefficients:
\begin{subequations}
\label{7}%
\begin{align}
\mathcal{R}_{m,n,\ell,s,t,u}^{\varepsilon(N)}  &  \equiv\frac{\left(
-1\right)  ^{u-t-2\delta_{\varepsilon e}}}{2^{N-\delta_{\varepsilon e}%
}i^{u-2\delta_{\varepsilon e}}}\left(
\begin{array}
[c]{c}%
N-u+\delta_{\varepsilon e}\\
s
\end{array}
\right)  \left(
\begin{array}
[c]{c}%
u-2\delta_{\varepsilon e}\\
t
\end{array}
\right)  \overline{\mathcal{B}}_{m-\delta_{\varepsilon e},n-\delta
_{\varepsilon e},\ell}^{(N-\delta_{\varepsilon e})}\text{,}\label{7a}\\
\mathcal{S}_{s,t,n}^{\varepsilon(N)}\left(  \wp,\tau\right)   &
=\frac{\left(  -1\right)  ^{\Upsilon\left(  v+\delta_{\varepsilon e}\right)
}}{\sqrt{2\pi}k\Delta r}\frac{\left(  \wp^{2}+\gamma^{2}\right)
^{1/2}\left\vert v+\delta_{\varepsilon e}\right\vert +\gamma}{\left(  \wp
^{2}+\gamma^{2}\right)  ^{3/2}}\left(  \frac{\wp}{\gamma+\left(  \gamma
^{2}+\wp^{2}\right)  ^{1/2}}\right)  ^{\left\vert v+\delta_{\varepsilon
e}\right\vert }, \label{7b}%
\end{align}
with $u=m+n-2\ell$, $v=2\left(  s+t\right)  -N$, and
\end{subequations}\end{widetext}
\begin{subequations}
\label{8}%
\begin{align}
\Upsilon(\tilde{\nu})  &  =\left\{
\begin{array}
[c]{ccc}%
0 & \text{for even/odd} & \tilde{\nu}\geq0\\
0 & \text{for even} & \tilde{\nu}<0\\
1 & \text{for odd} & \tilde{\nu}<0
\end{array}
\right.  ,\label{8a}\\
\gamma\left(  \tau\right)   &  =-\left(  2k\Delta r\right)  ^{-1}+i\sqrt
{n}\Lambda\left(  \tau\right)  . \label{8b}%
\end{align}
Therefore, from the analytical expressions for the Fourier transforms given by
Eq. (\ref{6}), we readily derive the atomic momentum distribution (\ref{2}).

To illustrate the role of the interaction parameter in the momentum
distribution function, in Fig. 2 we display the 2D momentum distribution in
the dimensionless space $\wp_{x}/\Lambda\times$ $\wp_{y}/\Lambda$, computed
for the interaction parameters (a) $\Lambda=5$ and (b) $\Lambda=20$. As
expected, the resolution of the distribution function becomes better as the
interaction parameter $\Lambda$ is increased \cite{HAS, FH, BB}. Moreover, the
components of transverse momentum acquired by the atoms are given by a
summation over the Fourier transforms $F_{m,n}(\wp,\phi)$, which, because of
their dependence on the $\operatorname*{e}\nolimits^{-i\rho\left[  \wp
\cos(\theta-\phi)-\sqrt{n}\Lambda\right]  }$ term (see Eq. (\ref{6})), each
yield a radial transverse momentum of the atoms $\wp=\sqrt{n}\Lambda$, with
$n\leq N$.

Another important feature visible in Fig. 2 is that the phase factor in the
prepared atomic state is responsible for the asymmetry of the distributions,
here favoring the probabilities on the first and second quadrant of $\wp
_{x}/\Lambda\times$ $\wp_{y}/\Lambda$. As shown below, this asymmetry of the
distribution is an important ingredient to achieve atomic lithography. Here we
stress that the necessary presence of the ground state in the atomic
superposition produces a great number of atoms with no significant deflection
(see detailed discussion in Ref. \cite{Outro-Paper}), causing the distribution
around the origin ($n=0$) to reach values considerably larger than those for
$n>0$. Therefore, to highlight the discrete pattern of peaks for $n>0$, which
corresponds to atoms that have indeed interacted with the cavities light, we
have cut off in Fig. 2 the distributions around the origin, for $W>2\times
10^{-3}$\ in Fig. 2(a) and $W>5\times10^{-4}$ in Fig. 2(b). Based on the same
reasoning, we have neglected the distribution around the origin for purposes
of lithography.

We also observe in Fig. 2 that, by increasing the interaction parameter
$\Lambda$ and consequently the transverse momentum $\wp$, the atoms are
scattered to a larger region of the momentum space, at the expense of
decreasing probabilities. For this reason, for the purpose of litography,
i.e., to concentrate the probability distribution around a desired spot, it is
better to use small values of $\Lambda$. Assuming that the atoms are measured
on a screen located at a distance $L$ from the cavities, the transverse
displacement associated with each radius is giving by $r_{n}=\sqrt{n}%
\Lambda\hbar kL/mv$, where $v$ is the longitudinal atomic velocity. With
$L\sim0.5$m and typical $v\sim500$m/s, we obtain in the microwave regime:
$r_{n}\sim\sqrt{n}\Lambda/10$, giving radii on the nanometer scale for an
interaction parameter $\Lambda\sim10$, that are separated by decreasing
distances $r_{n+1}-$ $r_{n}\sim\left(  \sqrt{n+1}-\sqrt{n}\right)  \Lambda/10$
nm between concentric radii. This scheme provides subwavelength resolution in
the nanometer scale using microwaves, for a wide range of photon number
demanding the field to be treated in a quantum way.

While the cross-cavity OSG setup can be applied to two-mode tomography
\cite{Outro-Paper}, this device was designed from the start for the purpose of
atomic lithography. After all, it seems quite reasonable to expect to be able
to control the 2D deflection of the atomic beam by manipulating the
cavity-mode states. Pursuing this initial goal, our protocol to achieve atomic
lithography follows precisely from the manipulation of the amplitudes and
phases of coherent $\left\vert \alpha\right\rangle $ or squeezed coherent
$S_{\xi}\left\vert \alpha\right\rangle =\left\vert \alpha_{\xi}\right\rangle $
states ($\xi=r\operatorname*{e}\nolimits^{i\varphi}$ standing for the squeeze
parameters, with $\xi=0$ for the coherent state) previously prepared in both
cavity modes. As we shall now show, this manipulation enables us to modulate
the atomic distribution by concentrating this function around a desired spot.
To this end, we resort to a map that associates the (real) transverse momentum
components $\wp_{x},\wp_{y}$ acquired by the atoms with the field states
prepared in the two cavities, $a$ and $b$, which must be confined to their
(abstract) momentum-quadrature components, i.e., $\alpha_{\xi}%
=\operatorname*{e}\nolimits^{i\varphi_{\alpha}}\left\vert \alpha_{\xi
}\right\vert $ and $\beta_{\xi^{\prime}}=\operatorname*{e}\nolimits^{i\varphi
_{\beta}}\left\vert \beta_{\xi^{\prime}}\right\vert $, with $\varphi_{\alpha
},\varphi_{\beta}=\pm\pi/2$, respectively. While the choice of phases defines
the quadrant in which the maximum of the atomic distribution is located:
$\alpha_{\xi}=i\left\vert \alpha_{\xi}\right\vert $ and $\beta_{\xi^{\prime}%
}=i\left\vert \beta_{\xi^{\prime}}\right\vert $ defining the first quadrant of
the space $\wp_{x}\times$ $\wp_{y}$, $\alpha_{\xi}=-i\left\vert \alpha_{\xi
}\right\vert $ and $\beta_{\xi^{\prime}}=i\left\vert \beta_{\xi^{\prime}%
}\right\vert $ defining the second quadrant and so on, the amplitudes
$\left\vert \alpha_{\xi}\right\vert $ and $\left\vert \beta_{\xi^{\prime}%
}\right\vert $, and consequently the mean values $\bar{\alpha}_{\xi
}=\left\langle \alpha_{\xi}\right\vert a^{\dagger}a\left\vert \alpha_{\xi
}\right\rangle $ and $\bar{\beta}_{\xi^{\prime}}=\left\langle \beta
_{\xi^{\prime}}\right\vert b^{\dagger}b\left\vert \beta_{\xi^{\prime}%
}\right\rangle $, define the average radius and angle of the maximum of the
atomic distribution. More specifically, we obtain the relations%

\end{subequations}
\begin{subequations}
\label{9}%
\begin{align}
\bar{\wp}  &  =\left(  \bar{\wp}_{x}+\bar{\wp}_{y}\right)  ^{1/2}%
\approx\Lambda\left(  \bar{\alpha}_{\xi}+\bar{\beta}_{\xi^{\prime}}\right)
^{1/2}\text{,}\label{9a}\\
\bar{\phi}  &  \approx\operatorname*{sign}(\varphi_{\alpha}%
)\operatorname*{sign}(\varphi_{\beta})\tan^{-1}\sqrt{\bar{\beta}_{\xi^{\prime
}}/\bar{\alpha}_{\xi}}+\pi\delta_{\varphi_{\alpha},-\left\vert \varphi_{\beta
}\right\vert }\text{.} \label{9b}%
\end{align}
The quantum nature of the fields reveals itself in the discrete peaks with
mean momentum $\sqrt{n}\Lambda$. Since the expectation value of $n$ is
approximately the average total number of photons in the cavities
$\left\langle n\right\rangle \approx\bar{\alpha}_{\xi}+\bar{\beta}%
_{\xi^{\prime}}$, we infer that $\bar{\wp}_{x}\approx\Lambda\bar{\alpha}_{\xi
}$ and $\bar{\wp}_{y}\approx\Lambda\bar{\beta}_{\xi^{\prime}}$, and
consequently Eqs. (\ref{9a}) and (\ref{9b}).

Apart from the manipulation of the cavity mode states, we must stress that the
phase factor appearing in the prepared atomic superposition $\left(
\left\vert g\right\rangle +\operatorname*{e}\nolimits^{i\varkappa}\left\vert
e\right\rangle \right)  /\sqrt{2}$ is another important ingredient for the
achievement of atomic litography. We have found that the choice $\varkappa
=\pi/2$ maximizes the distribution around the desired $\bar{\wp}$ and
$\bar{\phi}$, so it will be adopted in our illustration of the lithography process.

We begin by showing the effectiveness of the map in Eq. (\ref{9}) and by
discussing the resolution of the atomic beam deflection ---its sharpness
around the desired spot--- achieved when coherent or squeezed coherent states
are prepared in both cavity modes. We demonstrate that the more a coherent
state is squeezed in the momentum quadrature, the better the resolution
becomes. Furthermore, besides the need to confine the fields to their
momentum-quadrature components, their squeezing must also be done in the same
field quadrature, i.e., $\varphi=\pi$.

In Fig. 3(a) we present the momentum distribution following from the coherent
states $\alpha_{0}=\beta_{0}=3.54i$, with $\Lambda=4$. We clearly observe a
peak located around the desired values $\bar{\wp}=20$ and $\bar{\phi}=\pi/4$,
in excellent agreement with the values derived from Eq. (\ref{9}). A view from
above of this momentum distribution is also presented (again disregardeding
the corresponding probabilities around the center), which seems to be more
convenient for tomographic purposes.

In Fig. 3(b), the atomic momentum distribution resulting from a squeezed state
generated from $\alpha$ $=$ $\beta=5.77i$ and with squeezing factors
$r=r^{\prime}=0.5$ (other parameters being the same as in Fig. 3(a)), is
presented, exhibiting a higher resolution achieved around the same target
$\bar{\wp}=20$ and $\bar{\phi}=\pi/4$. Indeed a sharper peak of the momentum
distribution is located around the desired spot. The region of the
distribution function concentrating substantial probabilities around the
desired spot has decreased significantly.\textbf{ }By increasing further the
squeezing factors to $r=r^{\prime}=1$, and using $\alpha$ $=$ $\beta=9.06i$ to
keep $\bar{\wp}=20$ and $\bar{\phi}=\pi/4$, we observe in Fig. 3(c) that the
resolution of the distribution is further enhanced.

Next, we demonstrate\ how to manipulate the radial and angular degrees of
freedom of the atomic deflection. Once more assuming $\Lambda=4$ and squeezed
states generated from $\alpha$ $=$ $5.7i$ and $\beta$ $=$ $7.1i$, with
$r=r^{\prime}=1$, in Fig. 4(a) we present the distribution associated with the
target $\bar{\wp}=15$ and $\bar{\phi}=5\pi/18$, showing that smaller values of
the radii $\bar{\wp}$ may be achieved. Although values of $\bar{\wp}$ larger
than $20$ may also be accessed, we limited ourselves to $\bar{\wp}\leq$ $20$
because of the large computational demand to compute Eq. (\ref{2}). Finally,
in Fig. 4(b), we take the same parameters as in Fig. 4(a), but squeezed states
generated from $\alpha$ $=$ $-5.7i$ and $\beta$ $=$ $7.1i$, associated with
the rotated target $\bar{\wp}=15$ and $\bar{\phi}=13\pi/18$.

In conclusion, we have thus presented a full quantum mechanical scheme for
atomic lithography and demonstrated its effectiveness and tunability. We
stress that, differently from previous set-ups, the cavity set-up provides a
tunable lithographic scheme, in the sense that it is sufficient to tune the
intracavity field to monitor the deflection angle of the atomic beam. Then,
the cross-cavity allows to reach full two-dimensional control of the beam
deviation since each cavity offers control over one spatial degree of freedom.
In particular, mask-based techniques require designing a specific mask for
each atomic pattern --- the light-based scheme requires only to tune the
fields to create a new pattern. Practically, it may be used to design
two-dimensional microstructures. It is worth stressing that our aim is not to
compare the performance of our quantum scheme with semiclassical atomic
lithography, but to demonstrate the possibility of building effective
potentials from the radiation-matter interaction alone. The methods developed
above also enable the simultaneous tomography of two-mode states, by measuring
the 2D atomic momentum distribution \cite{Outro-Paper}. We finally observe
that the 2D cross-cavity OSG can also be used to generate Schr\"{o}dinger-cat
atomic states and entangled atomic states in positional space, a goal that we
will pursue at the next step.
\end{subequations}

\begin{flushleft}
\textbf{{\Large {Acknowledgements}}}

\end{flushleft}

The authors thanks V. S. Bagnato and B. Baseia for suggesting the theme, and
C. J. Villas-B\^{o}as and S. S. Mizrahi for enlightening discussions. We also
acknowledge the support from PRP/USP within the Research Support Center
Initiative (NAP Q-NANO) and FAPESP, CNPQ and CAPES, Brazilian agencies.

\begin{figure*}

\includegraphics[width=1.0\columnwidth]{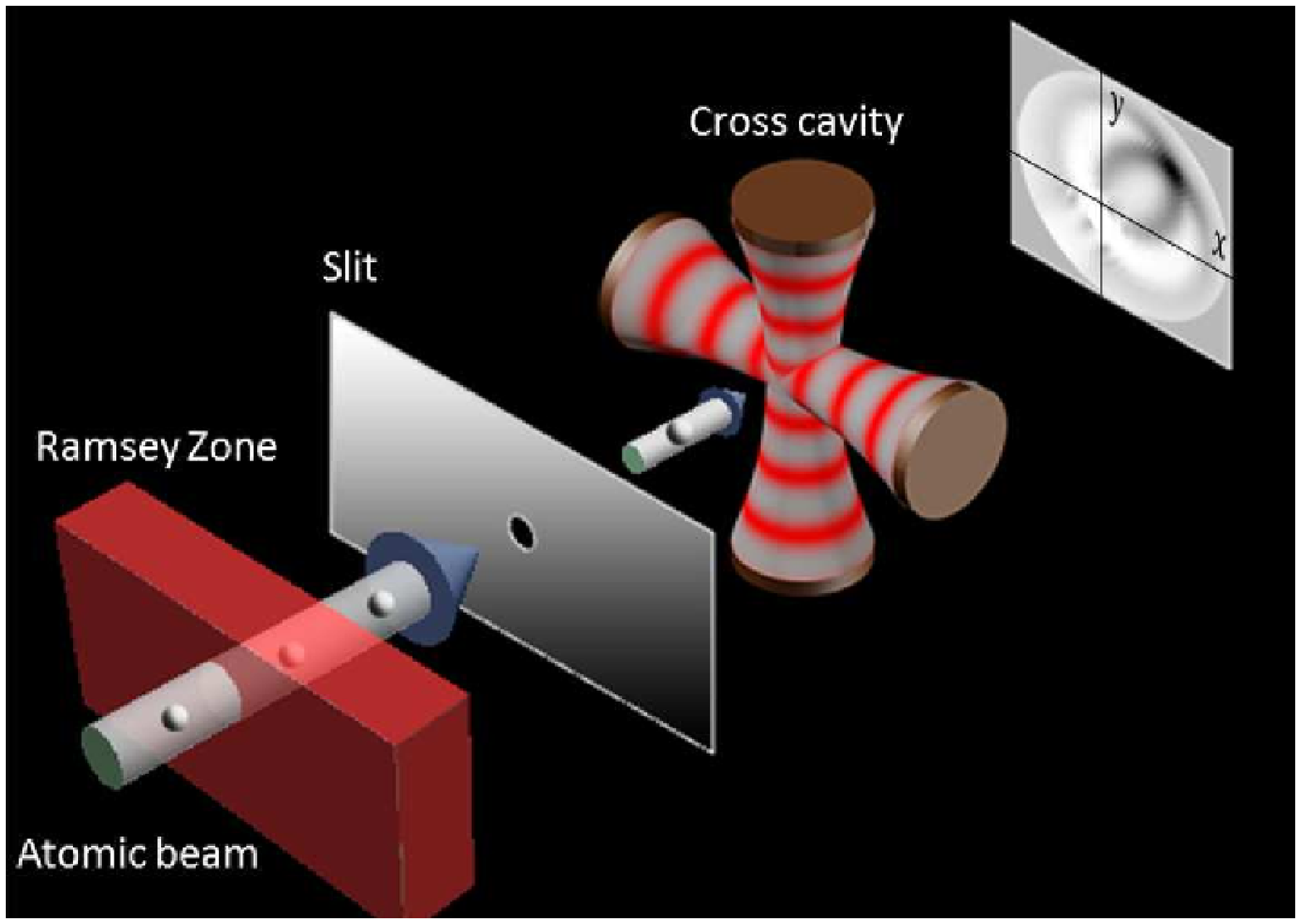}%
\caption{(color online). Sketch of the cross-cavity OSG.}%

\end{figure*}

\begin{figure*}

\includegraphics[width=1.0\columnwidth]{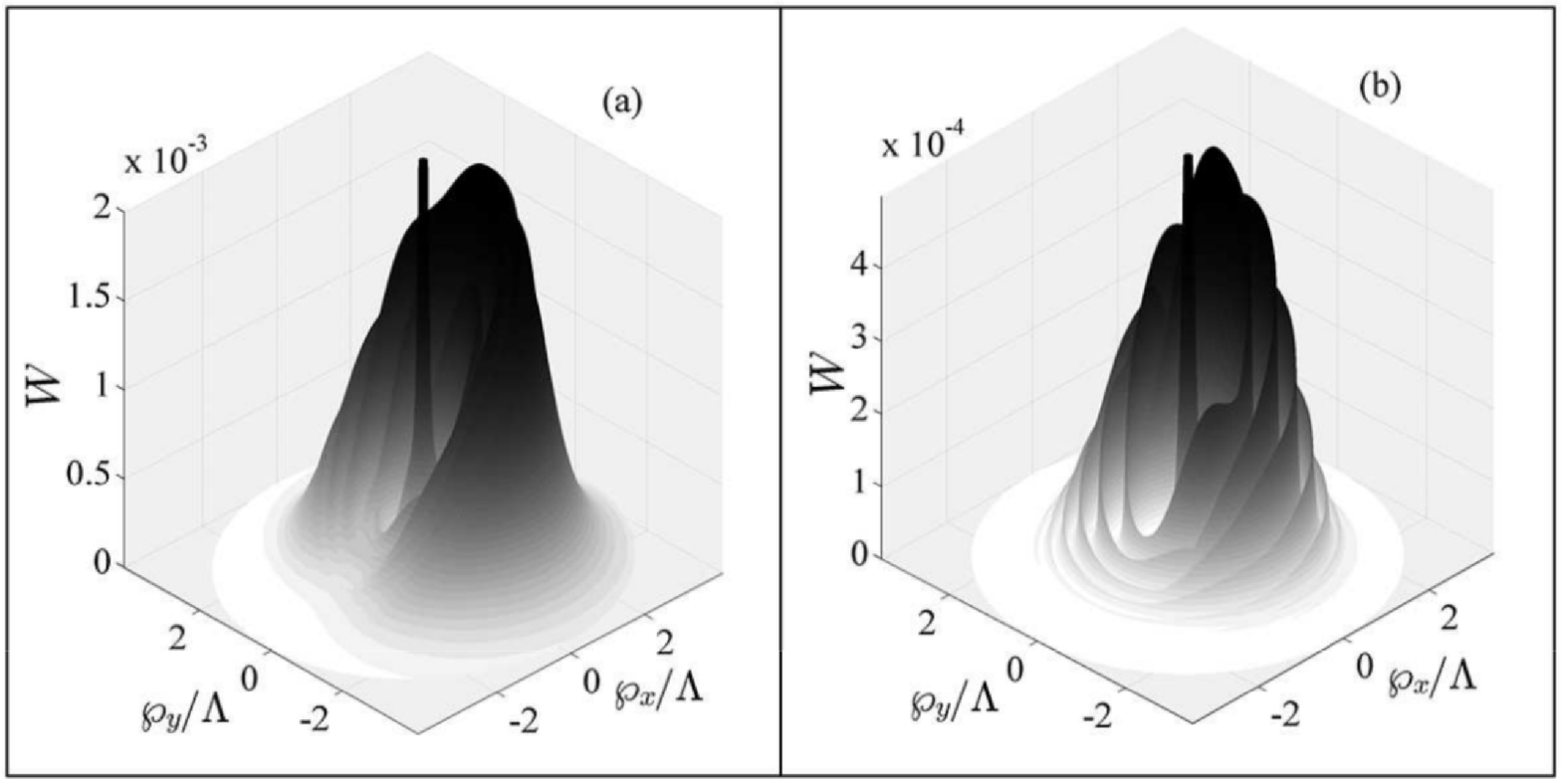}%
\caption{Atomic momentum distribution against $\wp_{x}/\Lambda\times$ $\wp
_{y}/\Lambda$ for (a) $\Lambda=5$ and (b) $\Lambda=20$. Simulations realized
for $k\Delta r=2\pi/10$, with the atoms initially prepared in the
superposition $\left(  \left\vert g\right\rangle +\operatorname*{e}%
\nolimits^{i\pi/3}\left\vert e\right\rangle \right)  /\sqrt{2}$, and the
cavity modes $a$ and $b$ in the product of coherent states $\left\vert
\psi_{field}\right\rangle =\left\vert \alpha\right\rangle \otimes\left\vert
\beta\right\rangle $, with $\alpha=\beta\operatorname*{e}\nolimits^{i\pi
/2}=1.5\operatorname*{e}\nolimits^{i\pi/2}$.}%

\end{figure*}

\begin{figure*}

\includegraphics[width=0.8\columnwidth]{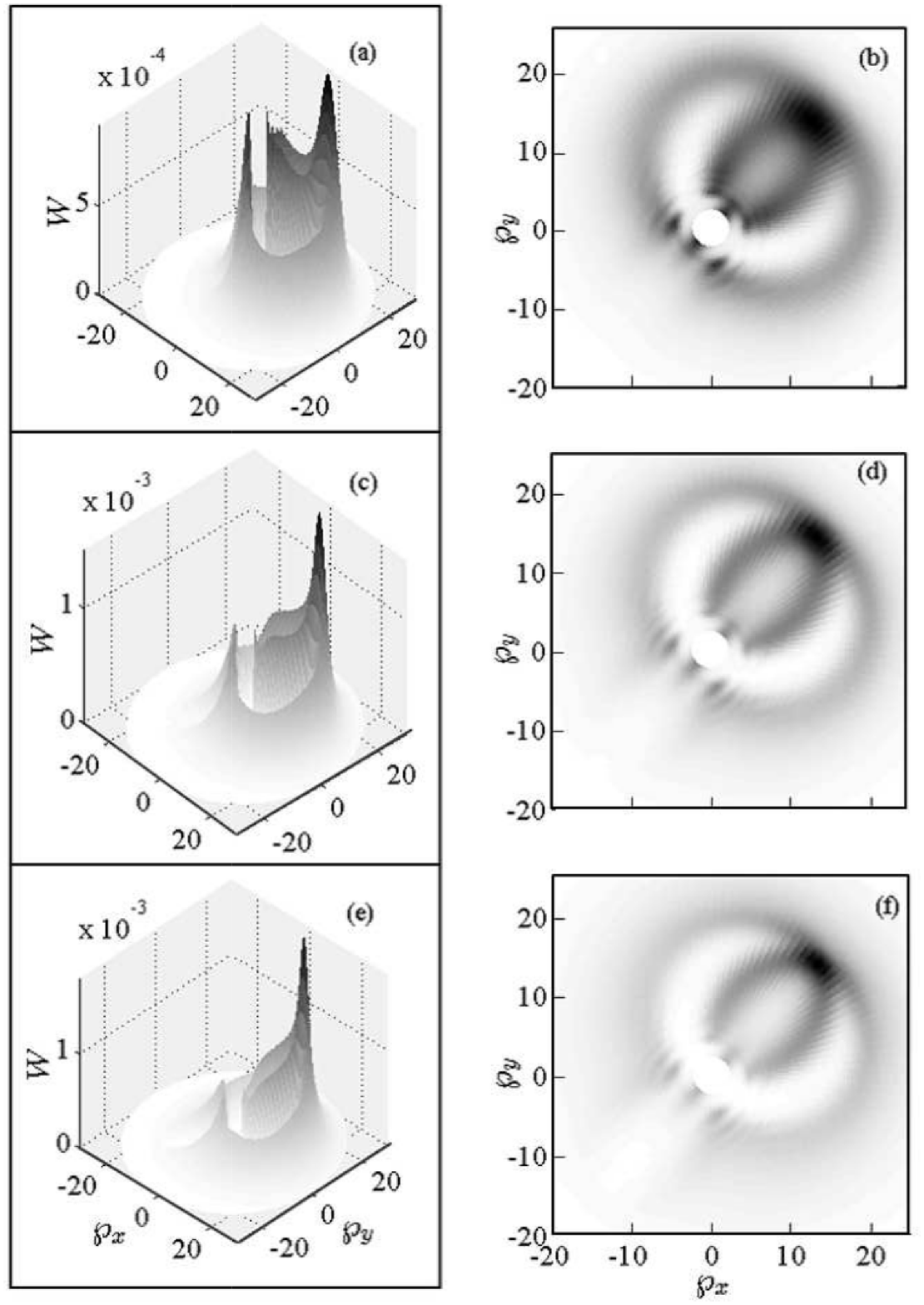}%
\caption{Atomic momentum distribution for $\Lambda=4$, $k\Delta r=2\pi/10$,
the atoms prepared in the superposition state $\left(  \left\vert
g\right\rangle +\operatorname*{e}\nolimits^{i\pi/2}\left\vert e\right\rangle
\right)  $, and the cavity modes in the (a) coherent states $\alpha_{0}%
=\beta_{0}=3.54i$, (b) squeezed coherent states with $\alpha$ $=$
$\beta=5.77i$ and squeezing factors $r=r^{\prime}=0.5$, and (c) squeezed
coherent states with $\alpha$ $=$ $\beta=9.06i$ and squeezing factors
$r=r^{\prime}=1$. In all three cases we aim at the target $\bar{\wp}=20$ and
$\bar{\phi}=\pi/4$.}%

\end{figure*}

\begin{figure*}

\includegraphics[width=0.8\columnwidth]{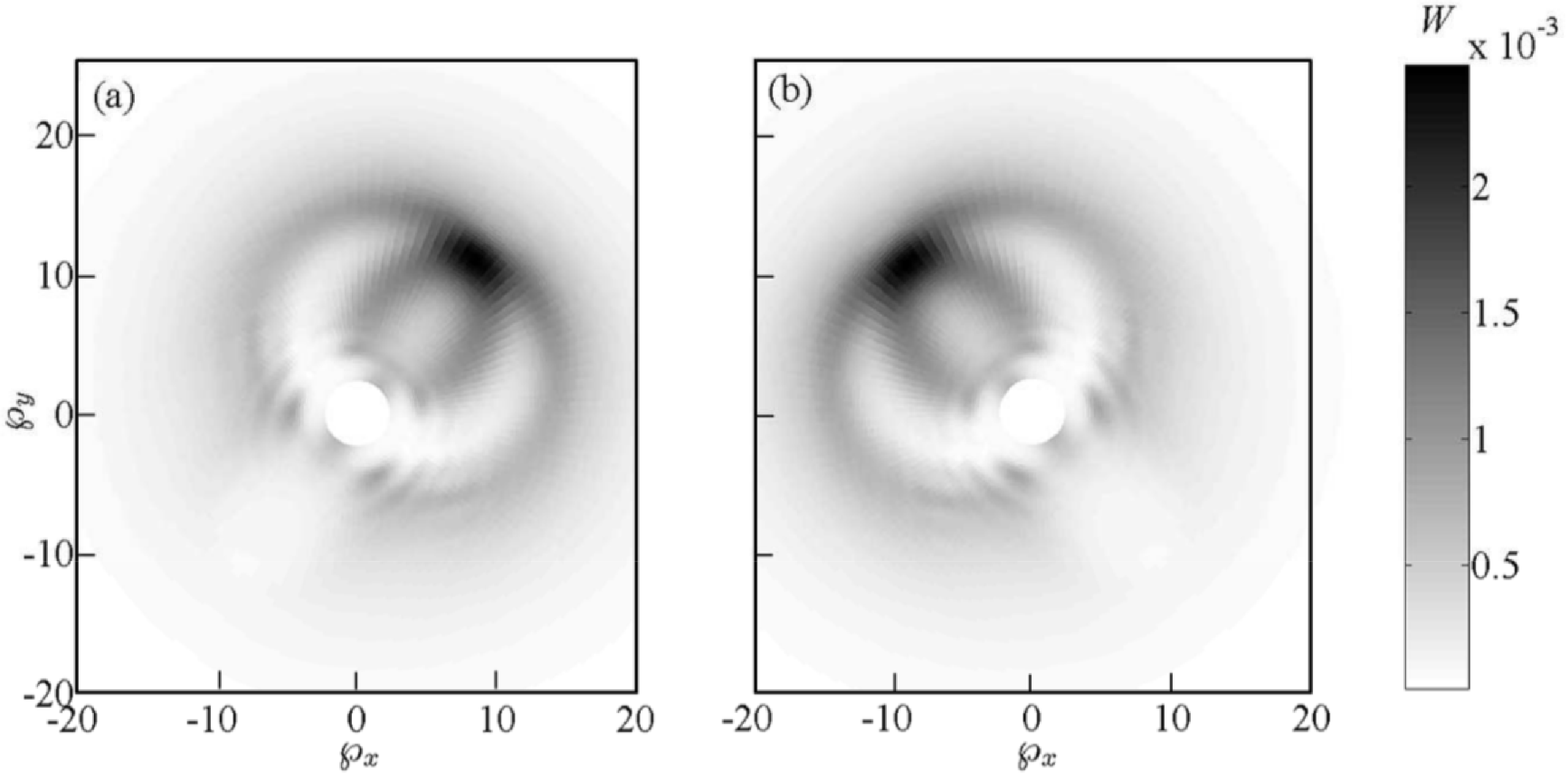}%
\caption{Atomic momentum distribution for $\Lambda=4$, $k\Delta r=2\pi/10$,
the atoms prepared in the superposition state $\left(  \left\vert
g\right\rangle +\operatorname*{e}\nolimits^{i\pi/2}\left\vert e\right\rangle
\right)  $, and the cavity modes in the squeezed states generated from the
squeezing factors $r=r^{\prime}=1$, with (a) $\alpha$ $=$ $5.7i$ and $\beta$
$=$ $7.1i$ and (b) $\alpha$ $=$ $-5.7i$ and $\beta$ $=$ $7.1i$.}%

\end{figure*}

\end{document}